\documentclass[11pt,fleqn]{article} 
 
\usepackage{a4} 
\usepackage{amstext} 
\usepackage{amsfonts} 
\usepackage{amssymb} 
\usepackage{color} 
\usepackage{epsfig} 
\usepackage{subfigure} 
\usepackage{verbatim} 
 
\newcommand{\gtapprox}{\raisebox{-0.5ex}{$\,\stackrel{>}{\scriptstyle\sim}\,$}}

\begin{document}


\begin{flushright} 
DESY 09-066 \\
HU-EP-09/17 \\ 
SFB/CPP-09-37
\end{flushright} 
 
\begin{center} 
 
{\huge \bf Comparing topological charge definitions}

{\huge \bf using topology fixing actions}
 
\vspace{0.7cm} 
 
\textbf{Falk~Bruckmann$^{1,2}$, Florian~Gruber$^1$, Karl~Jansen$^3$, Marina~Marinkovic$^{3,4,5}$, Carsten~Urbach$^4$, Marc~Wagner$^4$} 
 
$^1$Institut f\"ur Theoretische Physik, Universit\"at Regensburg, D-93040 Regensburg, Germany 
 
$^2$Institut f\"ur Theoretische Physik III, Universit\"at Erlangen, D-91058 Erlangen, Germany 
 
$^3$DESY, Platanenallee 6, D-15738 Zeuthen, Germany

$^4$Humboldt-Universit\"at zu Berlin, Institut f\"ur Physik, Newtonstra{\ss}e 15, D-12489 Berlin, Germany 
 
$^5$University of Belgrade, Faculty of Physics, Studentski trg 12, 11000 Belgrade, Serbia 
 
\vspace{0.7cm} 
 
 
\end{center} 
 
\vspace{0.1cm} 
 
\begin{tabular*}{16cm}{l@{\extracolsep{\fill}}r} \hline \end{tabular*} 
 
\vspace{-0.4cm} 
\begin{center} \textbf{Abstract} \end{center} 
\vspace{-0.4cm} 

We investigate both the hyperbolic action and the determinant ratio action designed to fix the topological charge on the lattice. We show to what extent topology is fixed depending on the parameters of these actions, keeping the physical situation fixed. At the same time the agreement between different definitions of topological charge -- the field theoretic and the index definition -- is directly correlated to the degree topology is fixed. Moreover, it turns out that the two definitions agree very well. We also study finite volume effects arising in the static potential and related quantities due to topology fixing.

\begin{tabular*}{16cm}{l@{\extracolsep{\fill}}r} \hline \end{tabular*} 
 
\thispagestyle{empty}


\newpage 
 
\setcounter{page}{1} 
 
\section{\label{SEC486}Introduction and motivation} 
 
Phenomena related to topology of gauge fields are of inherent non-perturbative nature. Using lattice techniques to investigate topology is, therefore, a natural choice. However, a lattice discretization leads to fundamental problems, e.g.\ naively discretizing the topological charge density yields non-integer values for the topological charge in general. Moreover, with different discretizations significantly different results can be encountered rendering it difficult to establish a unique concept of topology on the lattice.  
 
To stress the importance of topology, we mention 
the mass of the $\eta'$ meson, which originates to a large extent from  
topological effects.
In addition, the low lying Dirac eigenvalues are sensitive to the topological charge \cite{Leutwyler:1992yt}, 
which is also visible in random matrix theory \cite{Verbaarschot:2000dy} and  
in chiral perturbation theory in the so-called $\epsilon$-regime  
\cite{Hansen:1990un,Hansen:1990yg,Damgaard:2001js}.  
 
Therefore, it is highly  
desirable to have a well defined notion of topology on the lattice.  
One approach in this direction  
is to use the standard continuum expression of the topological charge density, $q \propto F \tilde{F}$, with a lattice discretization of the field strength, in particular improved versions of it. 
This is combined with suitable smearing techniques smoothing the sampled gauge field  
configurations \emph{a posteriori}.  
Another approach is to use  
the index of the overlap operator, which leads to a  
conceptually clean integer definition  
by means of the index theorem  
(but may also depend on parameters of the kernel of the overlap operator).  
While it has been demonstrated that smearing and other filtering methods  
agree to a large extent on the locations of topological structures  
\cite{bruckmann:06a,Ilgenfritz:2008ia,Bruckmann:2009vb},  
the agreement of the total topological charge is not satisfactory for precision measurements.  
 
A different approach to arrive at a unique definition of topological charge is to restrict the generation of lattice gauge configurations \emph{a priori} to those, which are smooth enough to resemble continuous gauge configurations. It has first been shown by L\"uscher \cite{luescher:82} that the topological charge has an  integer valued so-called geometric definition on lattice configurations, whose quantum fluctuations are bounded, i.e.\ whose plaquettes are close to the identity.  
In order to implement such a constraint on the plaquette values, modified gauge actions can be used such as the hyperbolic action of ref. \cite{luescher:98}. It has been demonstrated in practical simulations that the use of the hyperbolic action (and variants of it) indeed tends to fix topological charge \cite{fukaya:05,bietenholz:05}. 
 
Constraints on the plaquette that guarantee the locality of the overlap operator \cite{hernandez:98,neuberger:99a} represent similar admissibility conditions. 
Obeying these conditions assures that the kernel of the overlap operator has a spectral gap and hence the eigenvalue density of zero or near-zero eigenvalues vanishes. As a consequence, zero mode crossings and in turn changes of topological charge are avoided.
 
Such a method has been implemented in \cite{fukaya:06} with the aim 
to perform simulations with dynamical overlap fermions.  
The idea is to introduce a \emph{determinant ratio} employing  
Dirac operators at negative quark mass \cite{vranas:06}, which leads to   
the desired spectral gap and hence forbids topology changes. 
This is, because current dynamical overlap algorithms including approximations of the quark determinant 
make changes in the topological sector (i.e.\ overcoming the corresponding discontinuity) very costly, which results in a scaling with $O(V^2$) \cite{Schaefer:2006bk}. 
When performing simulations with fixed topological charge, there is no need to treat topological tunneling and simulations with dynamical overlap fermions become  
feasible as they are currently carried out by the JLQCD collaboration. For a recent review of these simulations see \cite{Hashimoto:2008fc}, where it is also demonstrated that topology is indeed fixed and that the method outlined above can be used in practice. 
 
While the simulations of the JLQCD collaboration aim at obtaining 
QCD results, which could  be compared to experiment, we want to address a more theoretical question. It is our goal to investigate, whether using a topology fixing action (either the hyperbolic action or the determinant ratio action) will lead to an agreement regarding the values of the topological charge, when  
different lattice definitions are used. Besides the theoretical and  
conceptual interest in this question a practically very useful  
consequence could be that much  
less computer time demanding definitions of the topological  
charge than the index of the overlap operator can be used.  
This would open the possibility to address topology on much larger lattices 
than it is presently feasible with the overlap operator. A first investigation concerning the agreement of different definitions of topological charge using the hyperbolic action has been performed in \cite{fukaya:05}.
 
In this work, we will concentrate on two definitions of topological charge, 
the field theoretic definition via an improved field strength tensor,  
which also involves some additional smearing,  
and the fermionic definition  
via the index of the overlap operator. 
As we will show below we find that using topology fixing actions 
goes hand in hand with  
a better agreement between these two lattice definitions of topological  
charge.  
 
 
It is important to note that working at a fixed value of the total topological charge in a finite volume violates cluster decomposition. In sufficiently large volumes local topological fluctuations still take place in every fixed topological sector. Therefore, fixing topology on the lattice to e.g. the trivial sector is supposed to provide correct physical results in the infinite volume limit with cluster decomposition becoming restored.  
We expect that fixing topology induces particular finite volume effects in accordance with \cite{brower:03,aoki:07}, where these effects have been addressed by using the $\theta$-dependence of physical quantities. 

Due to these finite volume effects physical observables receive different values, when evaluated in different topological sectors in finite (and small) volumes. 
In this paper we address the question by a preliminary investigation of the dependence of Wilson loops, 
the static potential and the Sommer parameter $r_0$. We indeed find clear evidence that 
such a dependence is present. 
We stress that it is
the original idea and long term plan of the present project to adress
topological finite size effects in a detailed manner computing many more observables 
than the ones studied here.
We consider this paper as a first and essential step in this
direction and we believe that it is an important basic work for
such a future investigation. Our main and novel result that for the
determinant ratio action different topological charge definitions agree,
allows to determine the topological charge also on very large
lattices where the overlap definition of topological charge becomes
impractical.

Our results indicate that the value of the topological charge will not
depend on the details of its definition in contrast to findings in the past on rather coarse configurations, without any smoothing.
We find this result quite remarkable and important.
The field theoretic definition is known to be strongly affected by local
lattice artefacts in general (and how they are treated e.g. under smearing).
With the determinant ratio action we employ a method that
constrains a global observable (the lowest eigenvalue of the kernel of
the overlap operator) almost by construction. However,
as we show for the first time here, it also leads to unique values of
topological charges through the field theoretic definition. 
 
The paper is organized as follows. In the next section we write down two definitions of topological charge on the lattice. The two sections afterwards are devoted to the implementation of two topology fixing actions, the hyperbolic action and the determinant ratio action. We compare obtained results to those from the standard Wilson action. In section~\ref{SEC005} we investigate the dependence of the static potential and related quantities on the topological sector. We close with a brief comparison and conclusions.


\section{\label{SEC001}Definitions of the topological charge} 
 
\subsection{Field theoretic definition} 
The straightforward definition of topological charge on the lattice is the discretization of the standard continuum expression, 
\begin{equation}
Q^{\rm f} \ \ = \ \ \frac{1}{32\pi^2}\sum_x \epsilon_{\mu\nu\rho\sigma}{\rm Tr}\Big(F_{\mu\nu}(x)F_{\rho\sigma}(x)\Big) ,
\label{topfield}
\end{equation}
where $F_{\mu\nu}(x)$ is some lattice field strength tensor consisting of appropriate linear combinations of Wilson loops. 
Instead of the naive discretization of the field strength tensor, which combines only plaquettes, i.e.\ the smallest possible Wilson loops $W_{(1,1)}$, we employ an improved field strength tensor \cite{bilson-thompson:02}
\begin{equation}
 F_{\mu\nu}(x) \ \ = \ \ k_1 C_{\mu\nu}^{(1\times1)}(x)+ k_2 C_{\mu\nu}^{(2\times2)}(x) + k_3 C_{\mu\nu}^{(3\times3)}(x),
\end{equation}
where $C_{\mu\nu}^{(n\times n)}(x)$ are the clover averages of $n\times n$ Wilson loops 
 attached to the lattice site $x$. In order to achieve $O(a^4)$ improvement at the tree level, we choose $k_1=1.5$, $k_2=-0.15$ and $k_3=1/90$. 


In addition, smoothing or filtering techniques are needed to come closer to integer values for $Q$, i.e. reducing renormalization factors originating from short-range fluctuations. We use 10 steps of APE smearing\footnote{Stout smearing \cite{Morningstar:2003gk} has been shown to lead to very similar results for the topological charge \cite{Bruckmann:2009vb}.}  
before evaluating eq.\ (\ref{topfield}).  
APE smearing \cite{Falcioni:1984ei,Albanese:1987ds} is an iterative procedure, where links are replaced by weighted averages of the old links $U_{\mu}$ and their attached staples $\tilde{U}_\mu^\nu(x)=U_{\nu}(x)U_{\mu}(x+\hat{\nu})U^{\dagger}_{\nu}(x+\hat{\mu})$ projected back to the gauge group: 
\begin{equation} 
U_\mu \ \ \to \ \ P_{SU(3)}\bigg\{(1-\alpha)U_\mu+\frac{\alpha}{6}\sum_{\nu\neq\pm\mu}\tilde{U}_\mu^\nu\bigg\} . 
\end{equation}  
The projection onto $SU(3)$ is not unique. $P_{SU(3)}(W)$ projects $W$ onto that $V \in SU(3)$ that maximizes $\textrm{Re}(\textrm{Tr}(V W^\dagger))$,
where the maximum is found iteratively as implemented in the CHROMA software for lattice QCD \cite{Edwards2005}.
 Setting the weight parameter $\alpha=0.45$, APE smearing has been argued to be equivalent to RG cycling \cite{DeGrand:1997ss}.

Smearing suppresses UV fluctuations in every observable, but is also biased towards classical solutions in the long run. Even in the short run it may destroy small topological objects. Hence, 10 steps can be considered as a compromise between obtaining topological charges closer to integer values and the destruction of topological objects.

\subsection{Definition via the index} 
 
An alternative and conceptually clean way of defining the topological  
charge motivated by the index theorem in the continuum is to use the index of the overlap operator  
\cite{Hasenfratz:1998ri,Niedermayer:1998bi}. 
The overlap operator, 
which is a solution of the Ginsparg-Wilson relation \cite{Ginsparg:1981bj}, has been introduced by Neuberger  
\cite{Neuberger:1997fp,Neuberger:1998wv} and reads 
for massless quarks 
\begin{equation}  \label{overlap} 
D_{\rm ov} \ \ = \ \ 1 + A / \sqrt{A^\dagger A} \quad , \quad A \ \ = \ \  D_{\rm W} - m_0 , 
\label{overlap_and_kernel}
\end{equation} 
where $D_{\rm W}$ is the Wilson Dirac operator and we have chosen the mass parameter $m_0=1.6$ (as in \cite{fukaya:06,Aoki:2008tq}).  
In the massless limit, which we consider here, the overlap operator (cf.\ eq.\ (\ref{overlap})) has exact zero modes with a definite chirality. This allows to define the topological charge as %
\begin{equation} 
Q^{\rm i} \ \ = \ \ \frac{1}{2} \sum_x {\rm Tr}\Big(\gamma_{5} D_{\rm ov}(x,x)\Big) \ \ = \ \ {\rm index}(D_{\rm ov}) \ \ = \ \ N_--N_+\: , 
\label{topindex} 
\end{equation} 
where $N_{-(+)}$ denotes the number of zero modes with negative (positive) chirality. To identify the zero modes we follow the techniques described in \cite{Giusti:2002sm,Chiarappa:2006hz}. 
 
\subsection{Comparison of topological charge definitions on standard Wilson gauge \\ configurations} 
 
Let us give an example for the difference of the resulting topological charges, when using the two definitions given above. We consider gauge configurations generated with the standard Wilson action  
\begin{equation}  \label{Wilact} 
S_{\rm W}[U] \ \ = \ \ \beta \sum_{P} S_{P}(U_{P}) \quad , \quad S_{P}(U_{P}) \ \ = \ \ 1 - U_{P} \quad , \quad  
U_{P} \ \ = \ \ \frac{1}{3} \,\textrm{Re}\Big(\textrm{Tr}(W_{(1,1)})\Big) , 
\end{equation} 
where $U_{P}$ denotes the plaquette variable and $\beta=6/g_0^2$.

\begin{figure}[htb] 
\subfigure 
{\includegraphics[width=0.35\linewidth,angle=90,angle=90,angle=90]{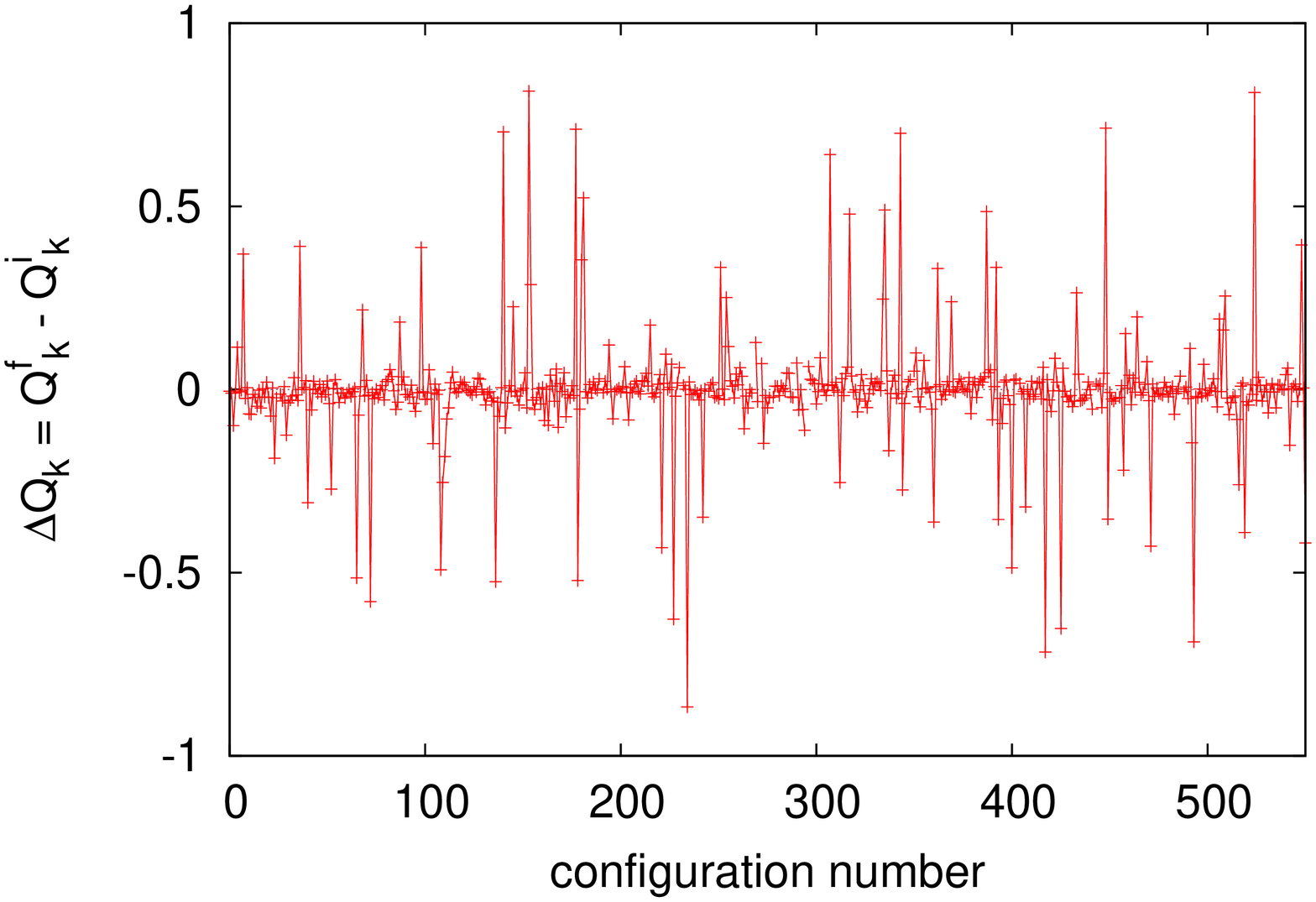}} \quad 
\subfigure 
{\includegraphics[width=0.35\linewidth,angle=90,angle=90,angle=90]{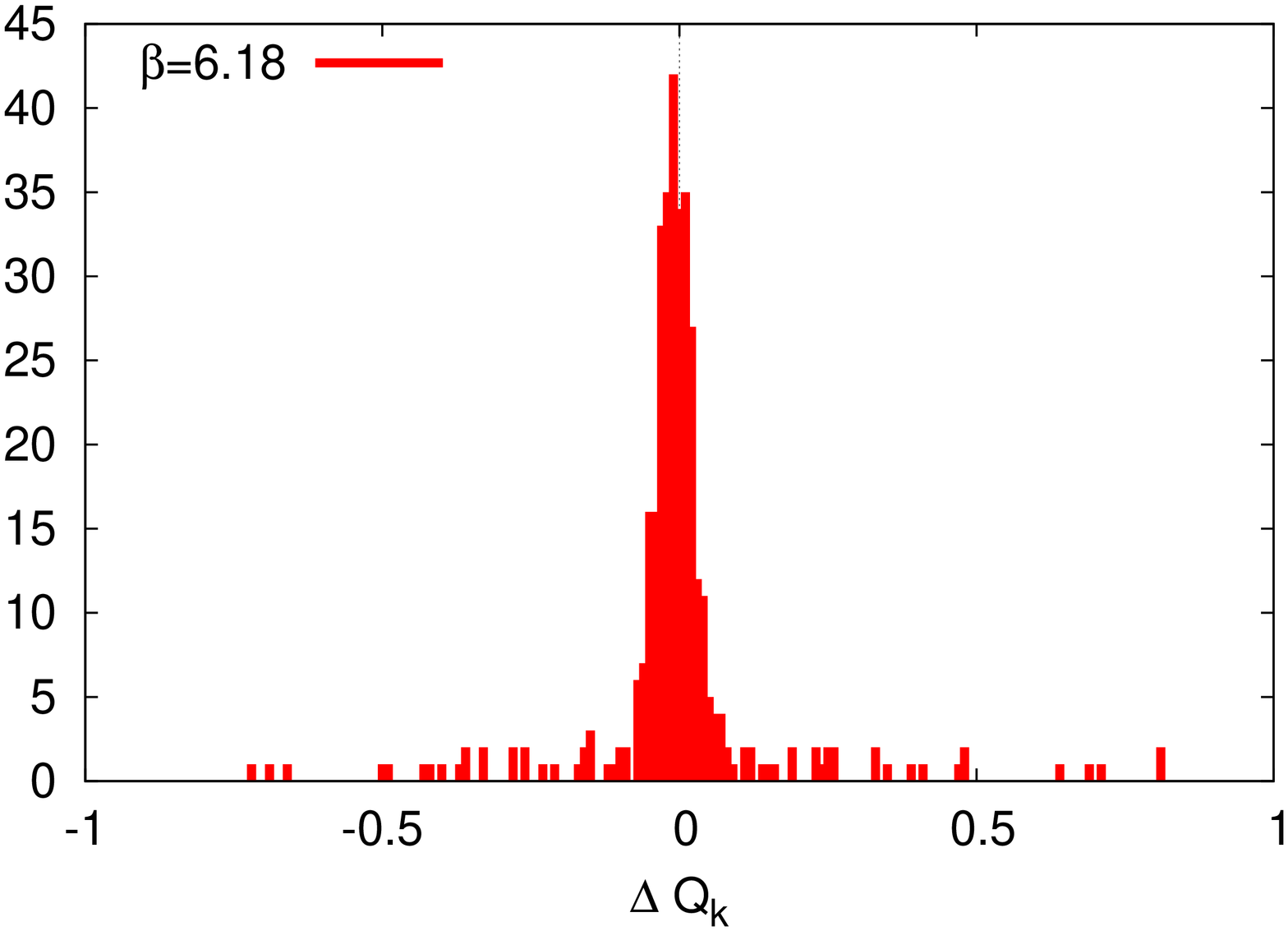}} 
\caption{The difference of the topological charge $\Delta Q_k = Q_{k}^{\rm f} - Q_{k}^{\rm i}$ as function of the configuration index $k$ (left) and the corresponding histogram (right) (Wilson plaquette gauge action).
}
\label{fig:wilson} 
\end{figure}

In the left panel of fig.~\ref{fig:wilson} we plot the Monte Carlo time history\footnote{One unit of Monte Carlo time corresponds to ten sweeps of link updating.} of the  
difference  
of the two topological charge definitions,
\begin{equation} 
\Delta Q_k \ \ = \ \ Q_{k}^{\rm f} - Q_{k}^{\rm i} ,
\label{deltak} 
\end{equation}  
as function of the configuration index $k$ using $\beta=6.18$ (this corresponds to $a \approx 0.07 \, \textrm{fm}$) and a lattice volume of $16^4$. We also show the corresponding histogram in the right panel of that figure.
As can clearly be seen frequent non-zero values of $\Delta Q_k$ are encountered. This shows that the two definitions of the topological charge yield different 
results. 
In particular, configurations with $|\Delta Q_k|>0.5$ would by rounding be ascribed to different topological sectors. 
Interestingly, we observed that most of these configurations are
associated with transitions in the index along the Monte Carlo time and
that during such transitions the field theoretical definition typically
recognizes the new topological charge not as quickly as the index
definition does.
 
Since we are already considering highly improved definitions of the topological charge, we do not follow the line of 
improving the  
topological charge or the filtering even further.  
 
An alternative route might be the application of improved {\em actions}.  
For example with the concept of perfect actions \cite{DeGrand:1995ji}
it can be expected that different definitions of the topological charge lead to coinciding values.  
 
Another possibility, which we will follow in this work, is to use  
actions that fix topology. Such actions have been tailored to address questions related to topology of gauge fields and give rise to the expectation that again  
better agreement of different topological charge definitions will be reached.  
Particular realizations of topology fixing actions are the hyperbolic action discussed 
in \cite{Fukaya:2003ph,bietenholz:05} and actions employing a determinant ratio 
of Wilson-type fermions \cite{fukaya:05,fukaya:06}. 
We will give the precise form of these actions in the next two sections. We will also discuss, whether using these actions indeed leads to a better agreement of the topological charge obtained from different definitions.

\section{\label{SEC002}The hyperbolic action} 
 
The hyperbolic action provides a cut-off for small plaquette values (i.e.\ large action density). It is given by 
\begin{equation}  \label{Shyp} 
S_{\epsilon,n}^{\rm hyp}[U] \ \ = \ \ \beta \sum_P S_{\epsilon,n,P}^{\rm hyp}(U_{P}) \quad , \quad S_{\epsilon,n,P}^{\rm hyp}(U_{P}) \ \ = \ \ \left\{ \begin{array}{cc} \frac{S_{P}(U_{P})}{(1 -  S_{P}(U_{P}) / \epsilon)^n} & \textrm{if }S_{P}(U_{P}) < \epsilon \\ \infty & \textrm{otherwise} \end{array} \right. , 
\end{equation} 
where for the rest of the paper we will use $n=1$. 
This form of the action has been suggested and used  
in \cite{Luscher:1998du,Luscher:1999un} 
for conceptual studies of chiral gauge theories on the lattice and it is 
motivated by the locality studies of the overlap operator \cite{hernandez:98}. There it has also been shown that for  
\begin{equation}  \label{epsineq} 
S_{P}(U_{P}) \ \ < \ \ \epsilon \ \ = \ \ \frac{2}{5 d (d-1)} \ \ = \ \ \frac{1}{30}
\end{equation} 
the Wilson kernel of the overlap  
operator has a spectral gap and hence zero mode crossings and, therefore,  
changes of topology are forbidden. Note that the bound in eq.\ (\ref{epsineq})  
has later been improved   
\cite{neuberger:99a} to  
\begin{equation}  \label{epsineqimp} 
S_{P}(U_{P}) \ \ < \ \ \epsilon \ \ = \ \ \frac{1}{(1 + 1/\sqrt{2}) d (d-1)} \ \ \simeq \ \ \frac{1}{20.5} . 
\end{equation} 
%
As we will describe below these theoretical bounds are typically not reached in practical  
simulations and, therefore, the fixing of topology will be imperfect. 
 
Note that the form of the action (cf.\ eq.\ (\ref{Shyp})) does not allow the 
application of a heatbath algorithm. Therefore, it has been  
simulated by using a local Hybrid Monte Carlo algorithm as proposed in  
\cite{Marenzoni:1993im} and discussed in its application to the  
present case in \cite{bietenholz:05}. 
 
The aim of our simulations is to use the hyperbolic action (and the determinant ratio action to be discussed later)  
\emph{in the same physical situation} (i.e.\ same physical volume and lattice spacing) defined by the example of the Wilson gauge action discussed in section~\ref{SEC001} and fig.~\ref{fig:wilson}.  
Consequently, when decreasing the value of $\epsilon$, also the value of $\beta$ has to be decreased in order to stay at the same physical lattice spacing as was demonstrated in \cite{bietenholz:05}.
%
  
In fact we took the same combinations of parameters  
$(\beta,1/\epsilon)$ detailed in \cite{bietenholz:05}, but performed independent simulations.  
The results for the plaquette expectation value  
$\langle U_P\rangle$, the integrated plaquette autocorrelation time $\tau_\textrm{int}^\textrm{plaq}$ and 
the acceptance rate and the step size of our local HMC algorithm are given  
in Table~\ref{table:Shyp}.   
In addition we list the frequency of topology changes $f_\textrm{top}$ in the corresponding  
runs, where we measured the topological charge through the index of the  
overlap operator. In the last column we give the static force parameter  
$r_0/a$ \cite{Sommer:1993ce} as computed from smeared Wilson loops using standard  
procedures \cite{Necco:2001xg}. From this the lattice spacing in physical units can be inferred, using $r_0 = 0.5 \, \textrm{fm}$. 
 
\begin{table}[htb] 
{\centering 
\begin{tabular}{|c|c|c|c|c|c|c|c|} 
\hline 
$\beta$ & $1/\epsilon$ & $\langle U_P\rangle$ & $\tau_\textrm{int}^\textrm{plaq}$ & acceptance rate & $d\tau$ & $f_\textrm{top}$ & $r_0/a$ \\ \hline \hline 
6.18 & 0.0 & 0.61185(2) & 8.2(9) & 0.998 & 0.1 & 0.0223 & 7.17(9) \\ \hline 
1.5 & 1.0 & 0.599858(5) & 2.2(2) & 0.997 & 0.1 & 0.0066 & 7.09(8) \\ \hline 
1.0 & 1.18 & 0.601518(3) & 1.20(6) & 0.995 & 0.1 & 0.0047 & 7.30(7) \\ \hline 
0.8 & 1.25 & 0.598366(3) & 1.13(5) & 0.992 & 0.1 & 0.0038 & 7.08(8) \\ \hline 
0.3 & 1.52 & 0.601035(2) & 0.82(2) & 0.877 & 0.1 & 0.0008 & 7.5(2) \\ 
\hline 
\end{tabular} 
\par} 
\centering 
\caption{Simulation parameters and results for the hyperbolic action (eq.~(\ref{Shyp}), lattice volume $16^4$). $\langle U_P\rangle$ denotes the plaquette expectation value, $\tau_\textrm{int}^\textrm{plaq}$ the corresponding integrated plaquette autocorrelation time, $d\tau$ the Monte Carlo step size, $f_\textrm{top}$ the frequency of topology changes and $r_0/a$ the static force parameter.} 
\label{table:Shyp} 
\end{table}

Table~\ref{table:Shyp} clearly shows that, when staying in the fixed physical situation of $r_0/a\approx 7$, which amounts to a lattice spacing of approximately $ 0.07 \, \textrm{fm}$, the frequency of topology flips is getting smaller, when decreasing $\epsilon$ as has already been noted in \cite{bietenholz:05}. Note, however, that even at our smallest value of $\epsilon=0.66$ ($1/\epsilon=1.52$ in Table~\ref{table:Shyp}) topology is not completely fixed. 
 
In fact at $r_0/a\approx 7$ it will not be possible  
to fix topology completely. The qualitative argument for this is the following: a roughly constant plaquette expectation value can be viewed as an indicator of being in a fixed physical situation. This is also suggested by Table~\ref{table:Shyp}: a value of $r_0/a\approx 7$ corresponds to $\langle U_P\rangle\approx0.6$. 
Decreasing $\epsilon$ beyond the desired value $\langle S_P\rangle$ is impossible, since according to eq.~(\ref{Shyp}) the weight of all configurations with action density $S_P > \epsilon$ becomes zero. 
Hence, there is a critical value
$\epsilon_{\rm crit}\approx \langle S_P\rangle$, which is the lowest value that can be reached 
for a given physical situation. For our present case $\epsilon$ cannot be lowered below $\epsilon_{\rm crit}\approx 1-0.6=0.4$.  
 
In Table~\ref{table:Shyp2} it is demonstrated that going to larger values of  
$r_0/a$ (i.e.\ to smaller values of the lattice spacing) the frequency of topology changes is reduced as expected.  
This indicates that at sufficiently small lattice spacings the hyperbolic action  
will lead to a complete fixing of topology. However, this will presumably  
happen at so small values of the lattice spacing that numerical simulations  
will be extremely demanding.

\begin{table}[htb] 
{\centering 
\begin{tabular}{|c|c|c|c|c|c|c|c|} 
\hline 
$\beta$ & $1/\epsilon$  & $d\tau$ & $f_{\rm top}$ & $r_0/a$ \\ \hline \hline 
0.25& 1.52 & 0.01 & 0.0041 & 6.7(2) \\ \hline 
0.3 & 1.52 & 0.01 & 0.0012 & 7.5(1)  \\ \hline 
0.1 & 1.64 & 0.01 & 0.0048 & 6.4(1) \\ \hline 
0.15& 1.64 & 0.01 & 0.0007 & 8.1(2) \\ 
\hline 
\end{tabular} 
\par} 
\centering 
\caption{Simulation parameters and results for the hyperbolic action of 
eq.~(\ref{Shyp}). Going to smaller values of the lattice spacing 
(i.e.\ to larger values of $r_0/a$) amounts to decreasing the frequency of topology changes $f_\textrm{top}$.} 
\label{table:Shyp2} 
\end{table} 

%
As a consequence of the above discussion our situation of $r_0/a\approx 7$ can only be achieved, when using values of $\epsilon \gtapprox 0.4$.
In practice, we stopped our simulation at $\epsilon = 1 / 1.52 \approx 0.66$, because for smaller values it is rather difficult to do a fine tuning of $\epsilon$ and/or $\beta$ to the desired physical situation. Naturally the question arises, whether for this value of $\epsilon$ there is already agreement between the topological charge values obtained from the two different definitions explained in section~\ref{SEC001}. As measure for the (dis)agreement we use 
\begin{equation} 
\Delta Q \ \ = \ \ \frac{1}{N_\textrm{conf}} \sum_{k=1}^{N_\textrm{conf}} |\Delta Q_k| , 
\label{deltaq} 
\end{equation}  
where the difference $\Delta Q_k$ is defined in eq.~(\ref{deltak}). 
 
\begin{figure}[htb] 
  \subfigure 
  {\includegraphics[width=0.35\linewidth,angle=90,angle=90,angle=90]{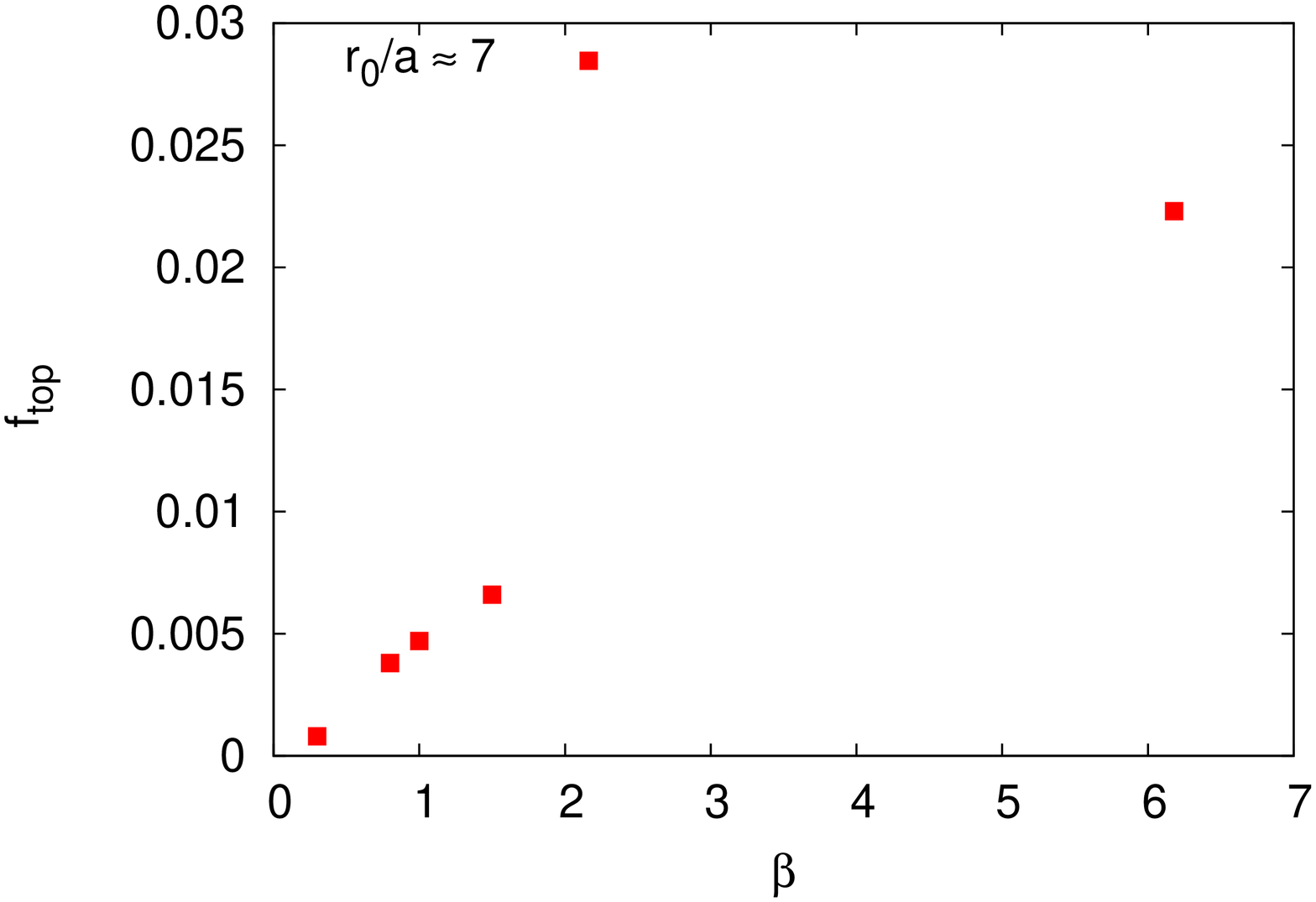}}\quad 
  \subfigure 
  {\includegraphics[width=0.35\linewidth,angle=90,angle=90,angle=90]{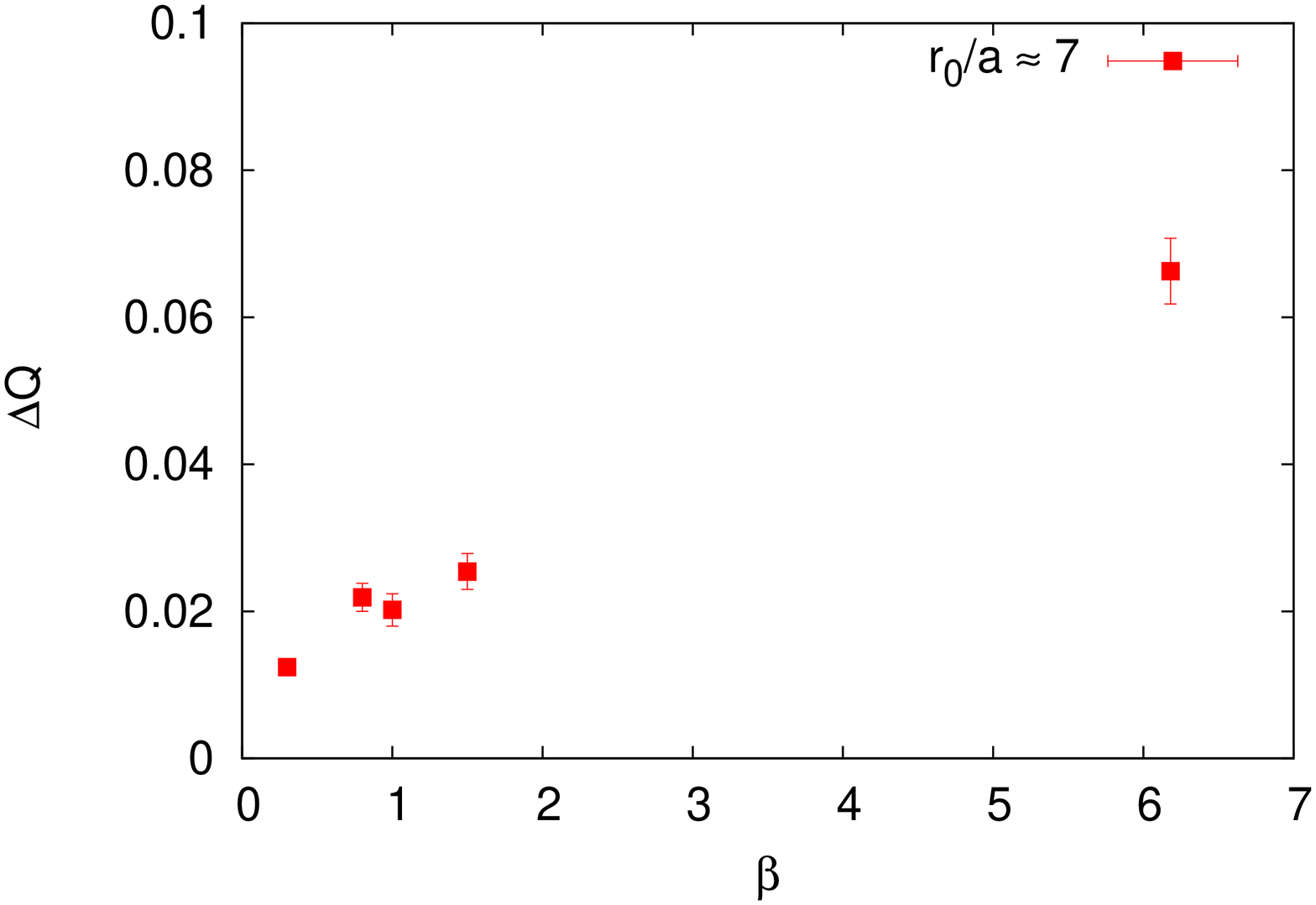}} 
 
\caption{The frequency of topology changes $f_\textrm{top}$ (left) and the difference of the topological charge $\Delta Q$ (right) as functions of $\beta$ for approximately fixed $r_0/a$ (hyperbolic action). Note that we do not show statistical errors for $f_\textrm{top}$, because of the long autocorrelation time of that quantity; nevertheless we consider the value of $f_\textrm{top}$ as an indicator to what extent topology is fixed.} 
\label{fig:ftophyp} 
\label{fig:lec} 
\end{figure}

In the left panel of fig.~\ref{fig:ftophyp} we show the frequency of topology changes and in the right panel of the same figure the values of $\Delta Q$ as functions of $\beta$.  
$\Delta Q$ exhibits a similar behavior as $f_{\rm top}$: it decreases as topology is fixed more and more. We conclude that topology  
fixing actions have indeed the potential to assure a unique results for the topological 
charge from different definitions. Remember, however, that at typical values of the lattice spacing it is not possible to fix topology sufficiently strongly with the hyperbolic action. Therefore, we consider an alternative approach using the determinant ratio action. 
 
\section{The determinant ratio action} 
 
The idea of fixing topology originates from the locality property of the overlap operator 
\cite{hernandez:98}. For a sufficiently fine lattice spacing ($a \lesssim 0.1 \, \textrm{fm}$) the kernel operator $A^2$ has a spectral gap and hence the overlap operator is local.  
This can be proven analytically using the theoretical bounds  
on $\epsilon$ for very small values of the lattice spacing. It has also been checked numerically.  
 
A spectral gap of the kernel operator in turn implies that there are no zero mode crossings and hence a change of topology (when using the index definition) is forbidden.  
This line of arguments suggests to include in the partition function a term  
\begin{equation} 
R_\mathrm{det} \ \ = \ \ \frac{\det\Big(A^\dagger A\Big)} 
{\det\Big(A^\dagger A+\mu_\textrm{tm}^2\Big)} .
\label{Sdet} 
\end{equation} 
Note that here the kernel operator $A$ (cf.\ eq.\ (\ref{overlap_and_kernel}))
is evaluated at negative quark mass $-m_0$ to guarantee the existence of a  
spectral gap. In the denominator in eq.~(\ref{Sdet}) a regulating twisted  
mass term $\mu_\textrm{tm}^2$ has been introduced, which controls the range of near zero modes that are to be suppressed.  
 
The determinant ratio action has been suggested in \cite{fukaya:05,fukaya:06}. There it has also been checked that it indeed fixes topology,  
when the topological charge is determined by the index of the overlap  
operator. The determinant ratio action has been extensively used in  
simulations with dynamical overlap fermions (see \cite{Hashimoto:2008fc}  
for a recent review) to avoid slowing down the algorithm at changes of the topological charge (see section~\ref{SEC486}). 
  
Here, we use the determinant ratio action as a local modification  
of the Wilson plaquette action. We do not include any dynamical fermions and, therefore, work in the quenched approximation. Our main motivation is to compare the field theoretic 
and the index definition of the topological charge and to determine to what  
extent they yield consistent results.   
 
The determinants that appear in eq.~(\ref{Sdet}) are treated by means of pseudo fermionic fields and a global Hybrid Monte Carlo algorithm \cite{Duane:1987de}. Details of this HMC algorithm can be found in \cite{Urbach:2005ji}. It uses even/odd preconditioning \cite{DeGrand:1990dk}, mass shift \cite{Hasenbusch:2001ne,Hasenbusch:2002ai} and multiple time step integrators. 
 
In a first step we have confirmed the results  
in \cite{fukaya:05,fukaya:06} that the topological charge,  
when computed via the index of the overlap operator, is indeed  
fixed. To this end we generated initial configurations on $16^4$ lattices  
with topological charge 
$Q^{\rm i}=0,1,2$ and performed simulations with the determinant ratio action. We have chosen  
$\beta= 6.063 $, $m_0= -1.6$ and $\mu_\textrm{tm}= 0.2$ (see below). We found that on samples of $300$ configurations separated by  
$10$ trajectories not a single topology change took place. 
Thus, topology seems to be fixed. 
 
To be able to perform a meaningful comparison with the hyperbolic action results from section~\ref{SEC002}, we tuned the value of $\beta$ such that $r_0/a \approx 7$. 
The results of this tuning are collected in Table~\ref{table:sdet}, where we 
list $\beta$, the average plaquette $\langle U_P\rangle$, the plaquette 
integrated autocorrelation time $\tau_\textrm{int}^\textrm{plaq}$, the acceptance rate 
and the force parameter $r_0/a$. Note that these results correspond to the $Q^{\rm i}=0$ sector. 
The main results of this table are that a value of $\beta=6.063$ leads  
to the desired scale of $r_0/a \approx 7$ and that the change in $\beta$ to reach this  
physical situation is much milder than in the case of the hyperbolic action.

\begin{table}[htb] 
{\centering 
\begin{tabular}{|c|c|c|c|c|} 
\hline 
$\beta$ & $\langle U_P\rangle$ & $\tau_\textrm{int}^\textrm{plaq}$ & acc.\ rate & $r_0/a$ \\ \hline \hline 
5.900 & 0.58850(3)& 3.7(6)& 0.85 & 5.43(4)\\ \hline 
6.000 & 0.59877(3)& 3.0(5)& 0.89 & 6.40(9)\\ \hline 
6.050 & 0.60358(3)& 2.9(5)& 0.90 & 6.69(6)\\ \hline 
6.063 & 0.60482(2)& 2.3(3)& 0.90 & 6.97(5)\\ \hline 
6.100 & 0.60830(3)& 2.5(4)& 0.91 & 8.04(22)\\ 
\hline 
\end{tabular} 
\par} 
\centering 
\caption{Simulation parameters and results for the determinant ratio action of eq.~(\ref{Sdet}).} 
\label{table:sdet} 
\end{table} 
 
\begin{figure}[htb] 
\begin{center} 
\includegraphics[width=0.35\linewidth,angle=90,angle=90,angle=90]{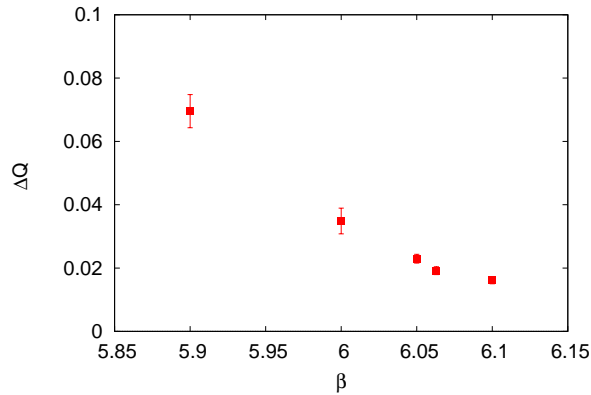} 
\caption{$\Delta Q$ as function of $\beta$ for the determinant ratio action. Note that contrary to fig.~\ref{fig:ftophyp} here we change the physical situation, i.e.\ the value of $r_0/a$ (cf.\ Table~\ref{table:sdet}).} 
\label{fig:deltaQdet} 
\end{center} 
\end{figure}

The crucial question is, of course, whether the two   
definitions of the topological  
charge, $Q^{\rm f}$ (cf.\ eq.\ (\ref{topfield})) and $Q^{\rm i}$ (cf.\ eq.\ (\ref{topindex})), agree. 
In fig.~\ref{fig:deltaQdet} the $\beta$ dependence of $\Delta Q$ (eq.~(\ref{deltaq}))  
is shown for the simulations detailed in Table~\ref{table:sdet}. Note that in all these runs $Q^{\rm i}=0$, i.e.\  
all simulations were performed in the topologically trivial sector. 
One can clearly see that $\Delta Q$ decreases, when approaching the continuum limit by increasing $\beta$. For $\beta=6.063$, i.e.\ for $r_0/a \approx 7$, $\Delta Q \approx 0.02$ 
indicates rather good agreement between the two definitions.  
 
\begin{figure}[htb] 
  \subfigure 
  {\includegraphics[width=0.35\linewidth,angle=90,angle=90,angle=90]{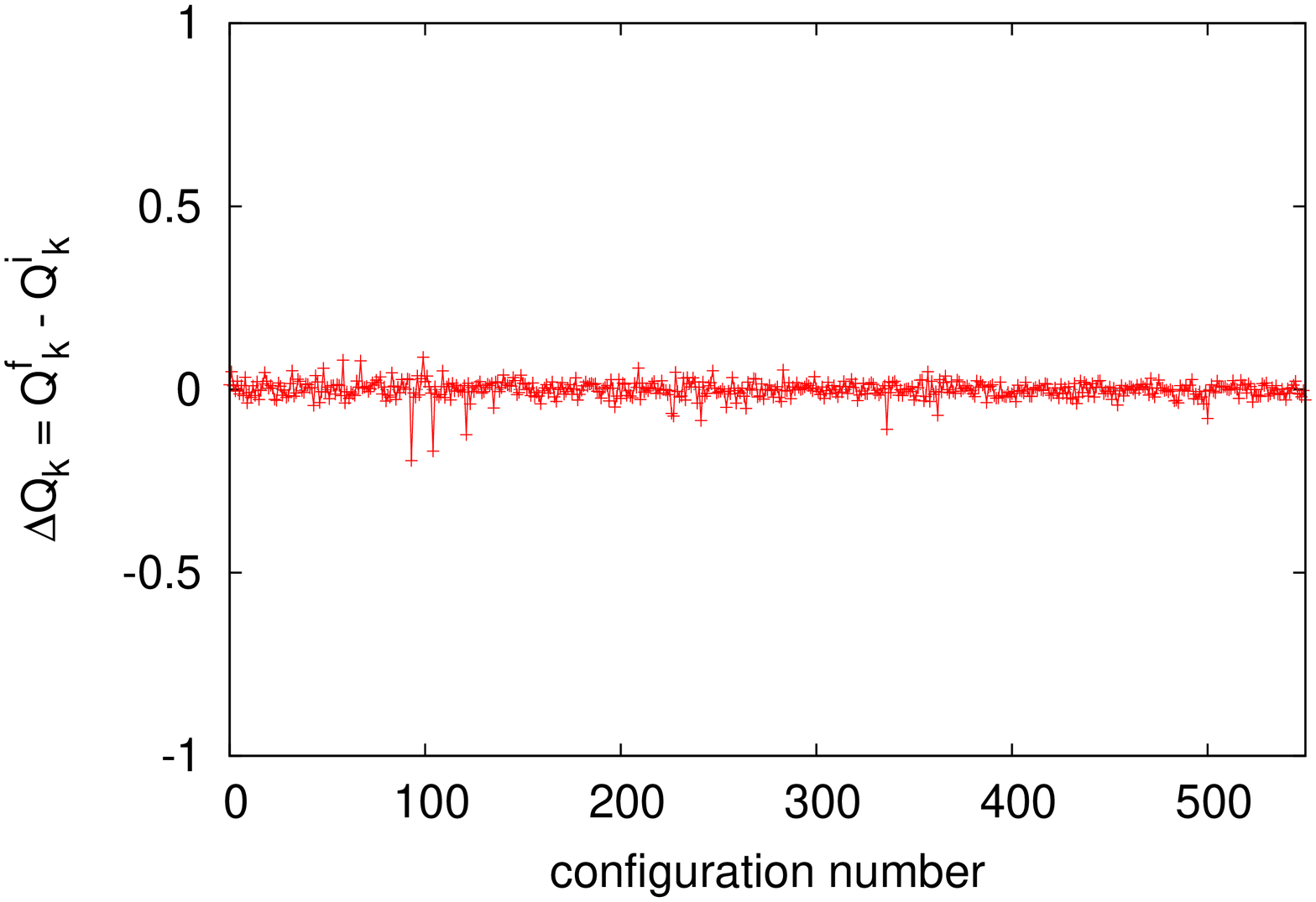}}\quad 
  \subfigure 
  {\includegraphics[width=0.35\linewidth,angle=90,angle=90,angle=90]{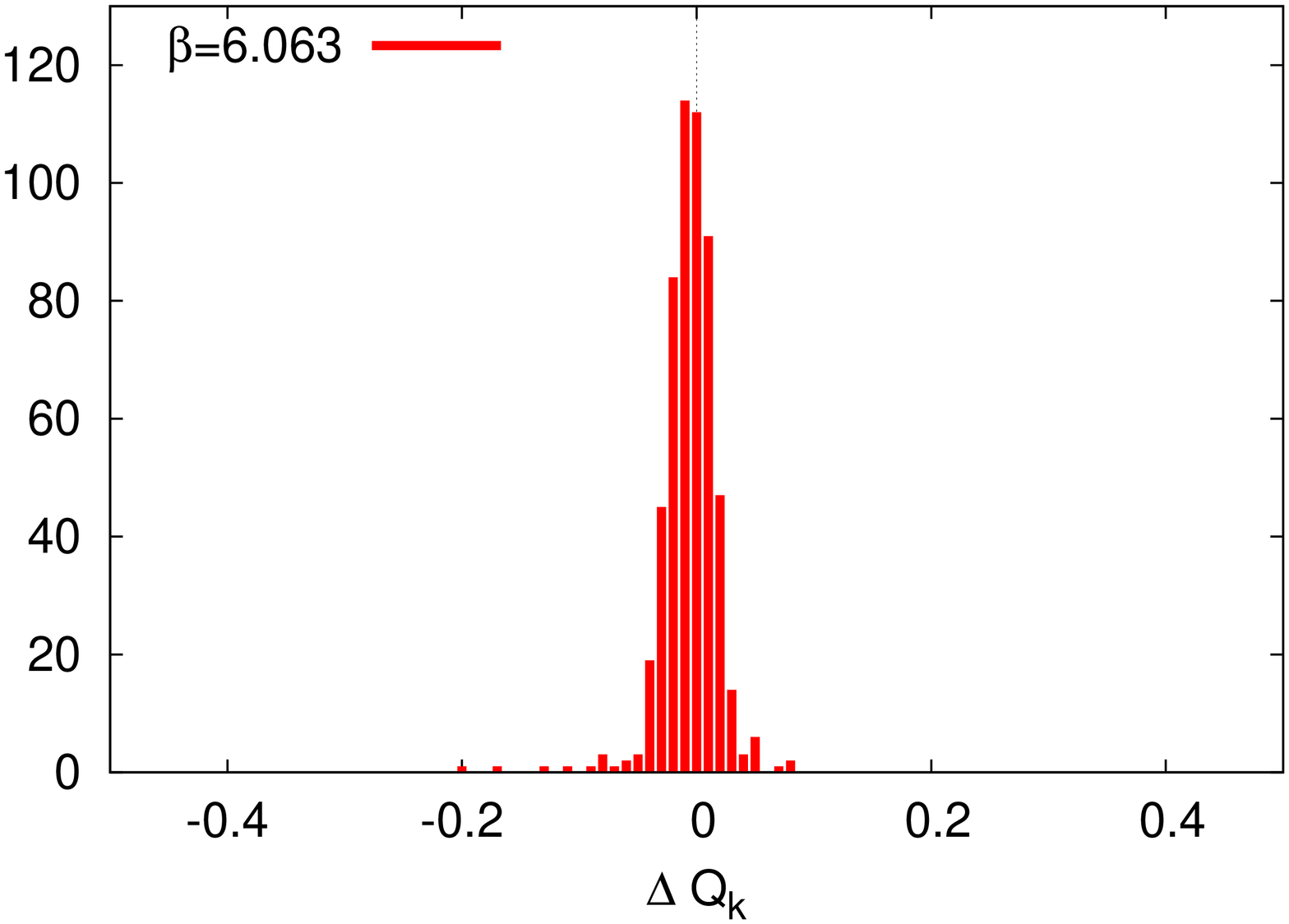}} 
  \subfigure  
  {\includegraphics[width=0.35\linewidth,angle=90,angle=90,angle=90]{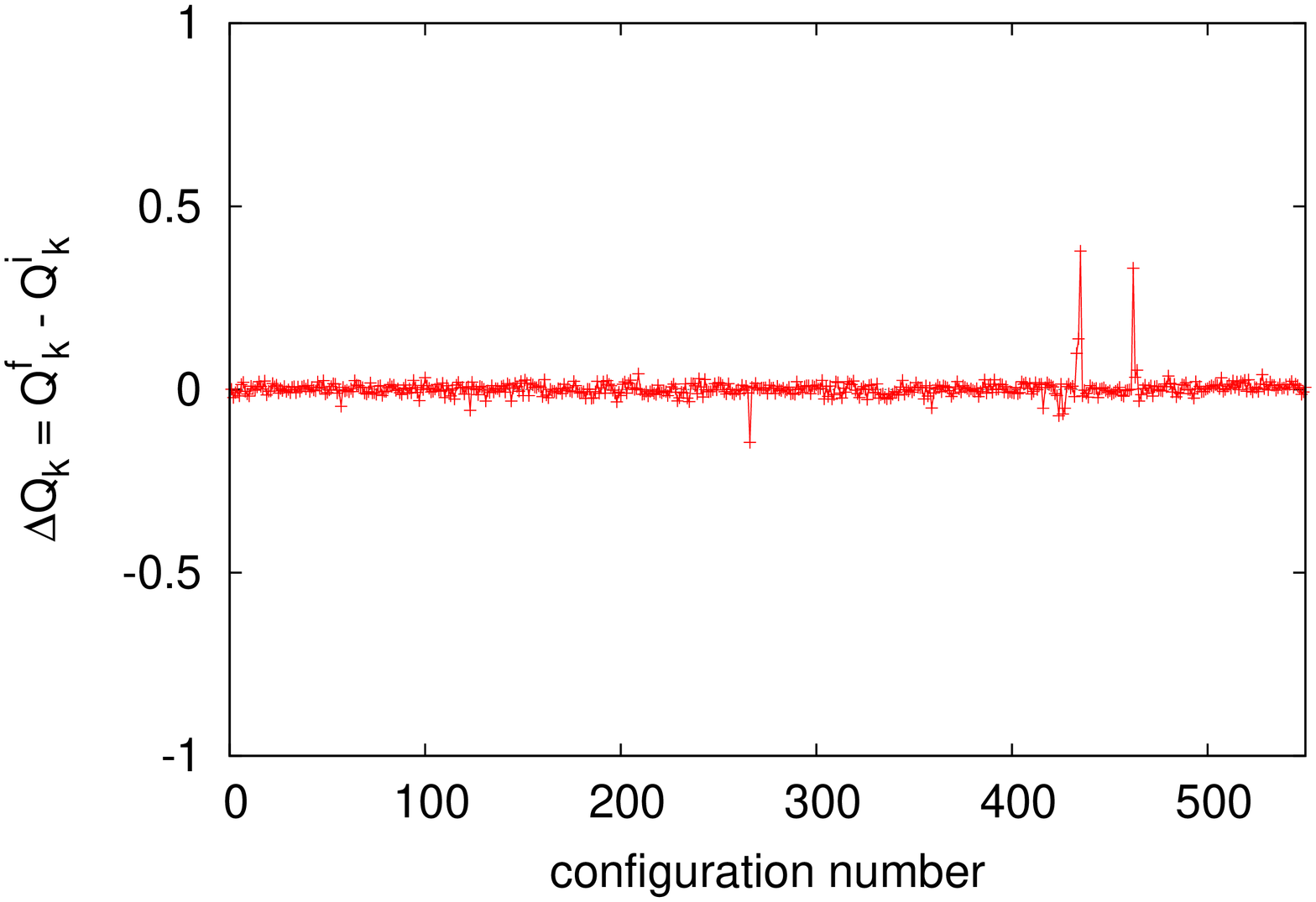}}\quad 
  \subfigure 
  {\includegraphics[width=0.35\linewidth,angle=90,angle=90,angle=90]{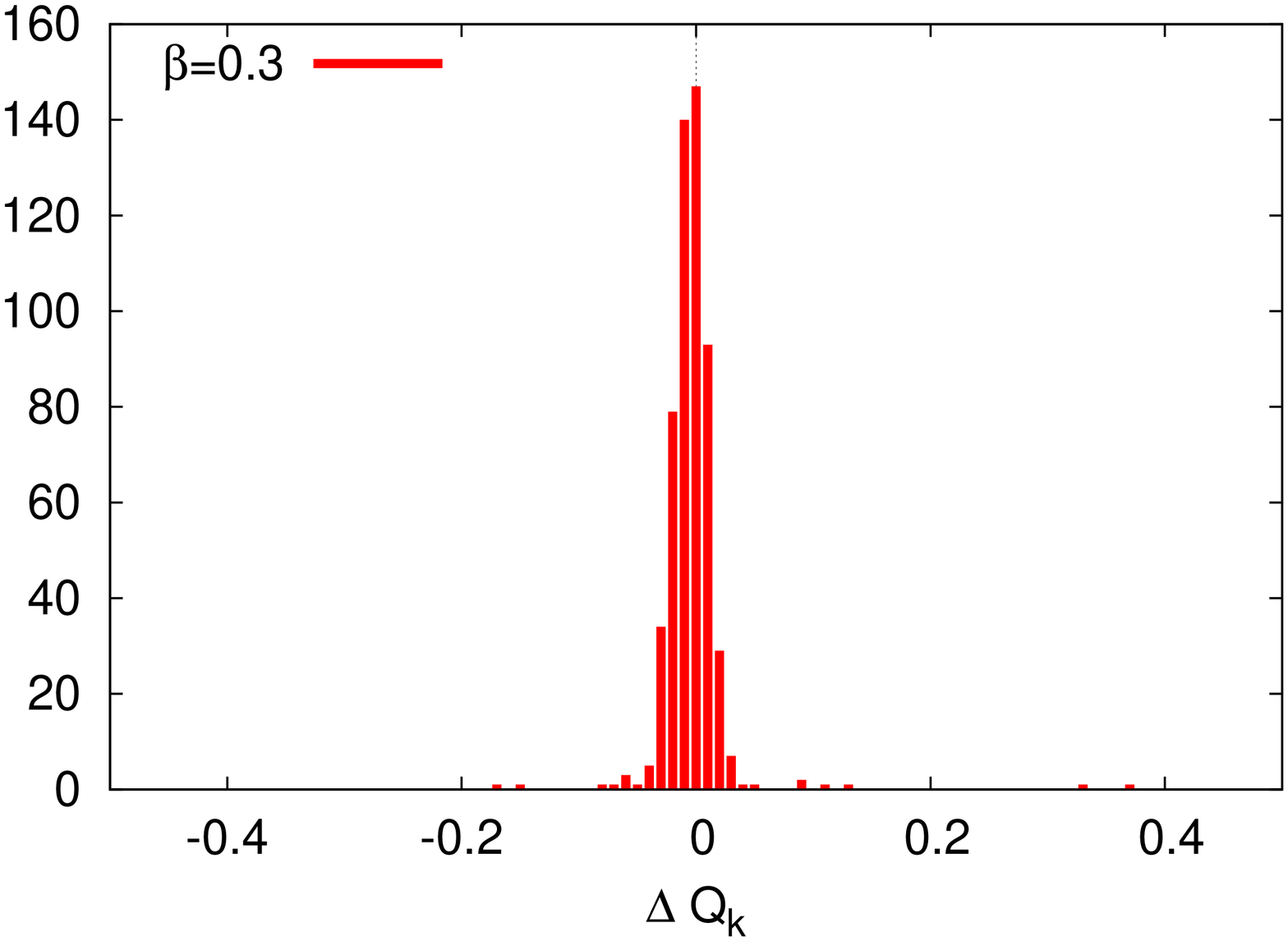}} 
\caption{The difference of the topological charge $\Delta Q_k = Q_{k}^{\rm f} - Q_{k}^{\rm i}$ as function of the  
configuration index $k$ (left) and the corresponding histogram (right) for the determinant ratio action at $\beta = 6.063$ (upper panels) and for the hyperbolic action at $\beta = 0.3$, $1/\epsilon=1.52$ (lower panels).   
} 
\label{detresults} 
\end{figure}

In the upper panels of fig.~\ref{detresults} we show the Monte Carlo history of $\Delta Q_k$, the disagreement of the two definitions, at $\beta=6.063$. When rounding the field theoretic results to the nearest integer, there is agreement for almost all gauge configurations. Only for the lower values of $\beta$ we have encountered in rare cases differences bigger than $0.5$. Corresponding plots for the hyperbolic action ($\beta = 0.3$, $1 / \epsilon = 1.52$) look essentially identical (cf.\ the lower panels of fig.~\ref{detresults}).
 
\begin{figure}[htb] 
\begin{center} 
{\includegraphics[width=0.35\linewidth,angle=90,angle=90,angle=90]{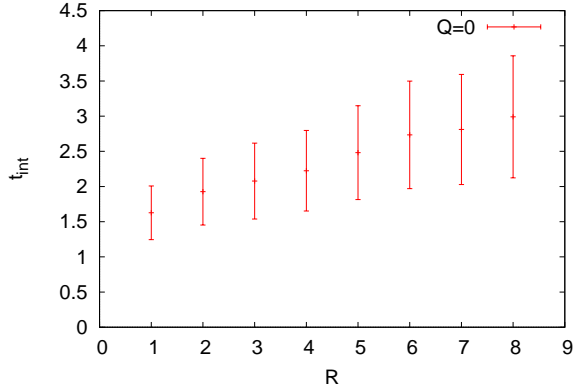}} 
\caption{Autocorrelation times of smeared Wilson loops with extension $R \times T$ as a functions of $R$ with $T=7$ fixed, here shown for a fixed 
topological sector, $Q=0$ 
(determinant ratio action, $\beta=6.063$, lattice volume $16^4$). Note that the autocorrelation time
in fig.~\ref{fig:wilsonloop}  is measured in units of configurations and each configuration is
separated by ten Monte Carlo iterations.} 
\label{fig:wilsonloop} 
\end{center} 
\end{figure}

We close this section with a short remark about autocorrelation times. As can be seen in Table~\ref{table:sdet} the autocorrelation times of the plaquette turn out to be rather short in our simulations.   
However, it is important to realize that the plaquette does not provide the  
largest autocorrelation time of the system. This can be seen from fig.~\ref{fig:wilsonloop}, where we show the autocorrelation times of smeared Wilson loops (extension $R \times T$) as functions of  
$R$ keeping $T=7$ fixed. 
Note that the autocorrelation time
in fig.~\ref{fig:wilsonloop}, $t_{int}$, is measured in units of configurations and each configuration is
separated by ten Monte Carlo iterations. Therefore the autocorrelation times
of the Wilson loops in fig.~\ref{fig:wilsonloop} are an order of magnitude larger than the
ones for the plaquette given in Table~\ref{table:sdet}.
Clearly, autocorrelation times become larger, when increasing $R$ until at $R \approx 6$ they level off and reach plateaus.  The increase of the smeared Wilson loops autocorrelation times is not a property characteristic for the topology fixing actions only \cite{Lippert:1997qy}.
In particular, when computing the static potential and from that $r_0 / a$ one should use these larger autocorrelation times, when computing statistical errors. 
 
\section{\label{SEC005}Dependence of the static potential and related quantities on \\ the topological sector}

Another important issue, which we intend to address in more detail in a subsequent publication, is the dependence of physical observables on topology, in particular the finite size effects in a given topological sector. Such results could be confronted with the theoretical expectations of \cite{brower:03,aoki:07}. 

\begin{figure}[htb] 
\subfigure{\includegraphics[width=0.35\linewidth,angle=90,angle=90,angle=90]{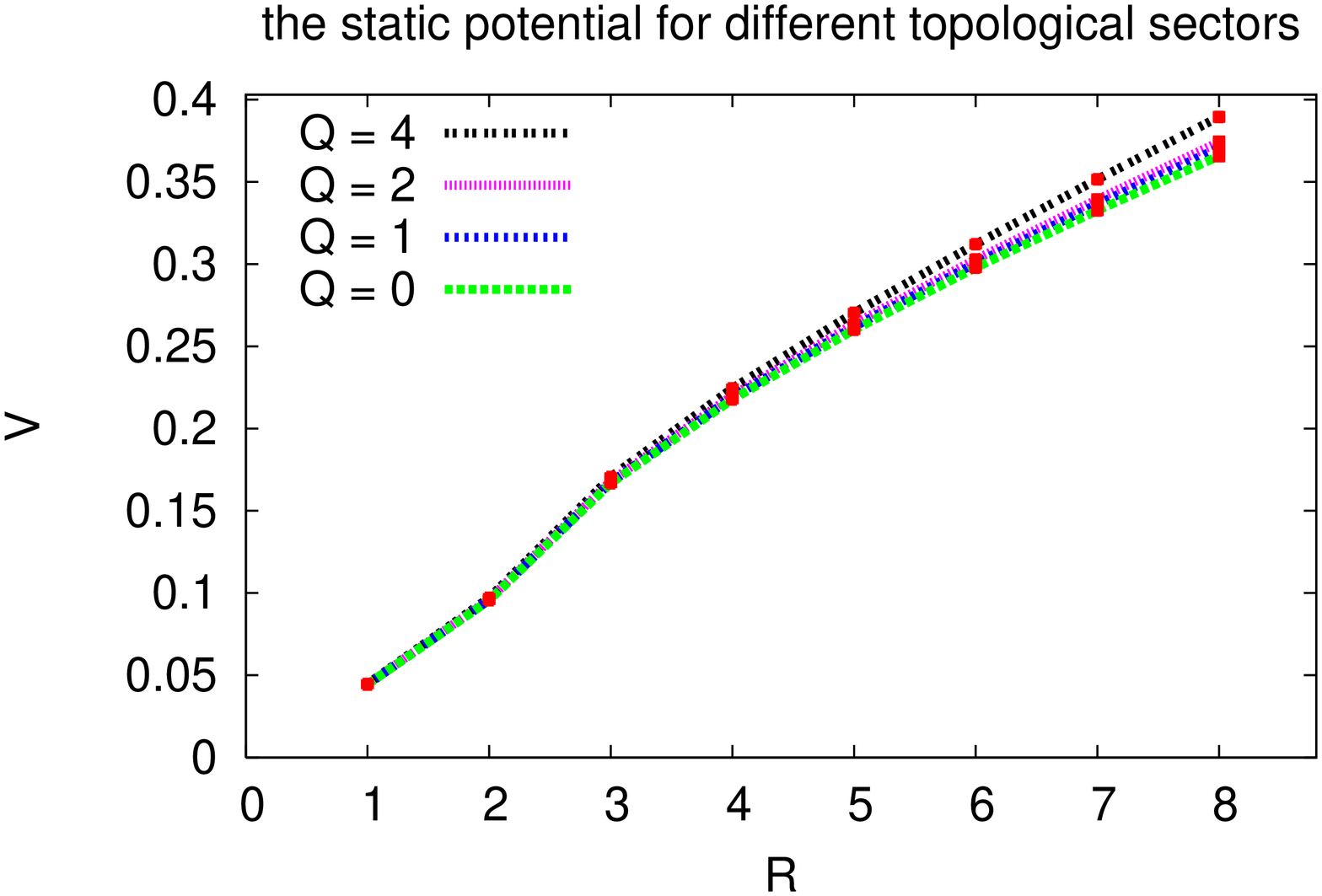}}\quad 
\subfigure{\includegraphics[width=0.35\linewidth,angle=90,angle=90,angle=90]{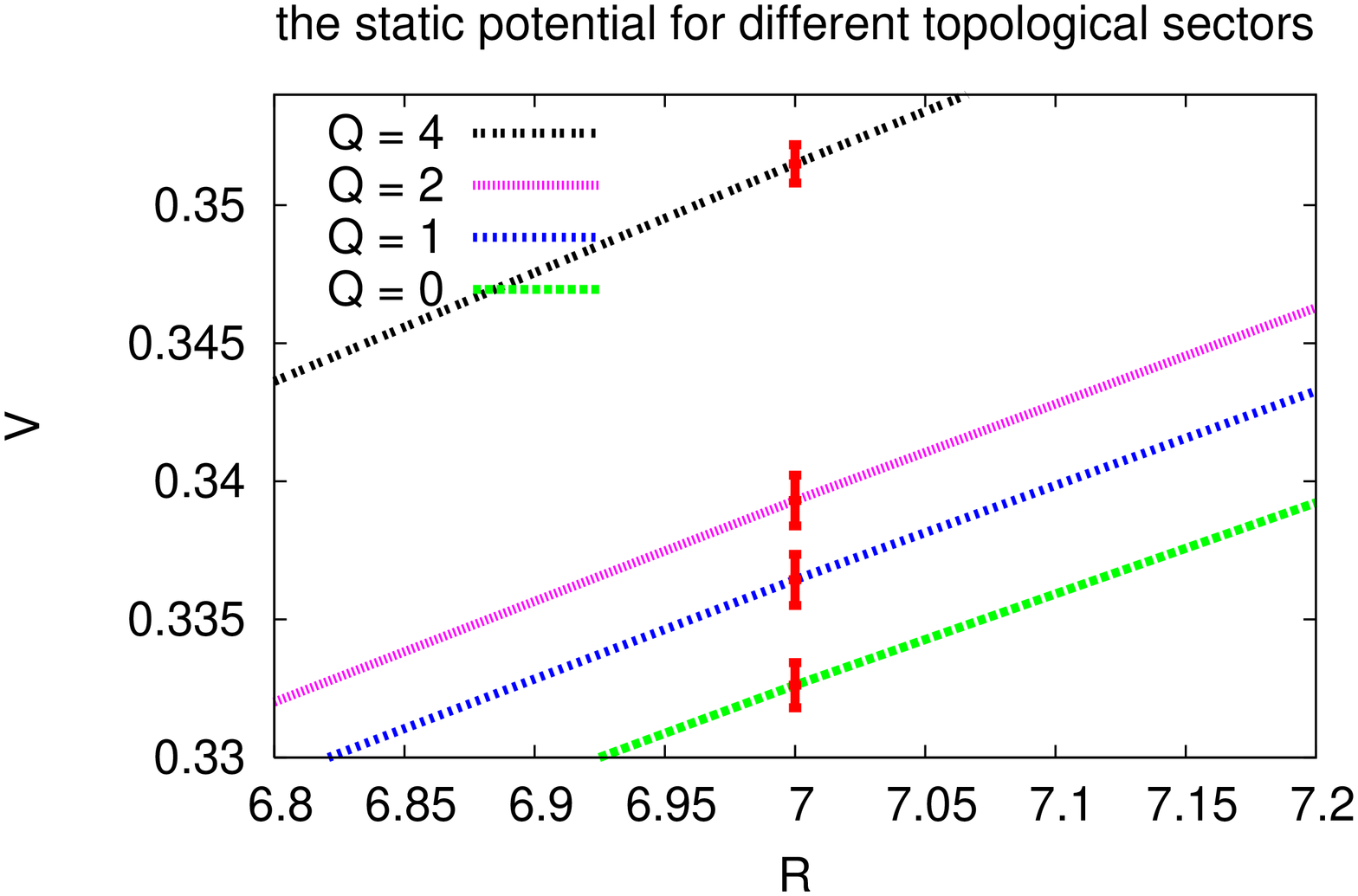}}

\caption{\label{FIG001}The static potential as a function of the separation for the topological sectors $Q^{\rm i} = \{0 \, , \, 1 \, , \, 2 \, , \, 4\}$ (left), and a zoomed plot of the same results (right).} 
\end{figure}

Here we study in a first step the $Q$ dependence of the static potential and related quantities. We employed the determinant ratio action at $\beta = 6.063$ using $16^4$ lattices. To identify the topological charge we used the index definition $Q^{\rm i}$. We computed smeared Wilson loop averages $\langle  W_{(R,T)} \rangle$ (extension $R \times T$), the static potential $V(R)$ at separation $R$ and the Sommer parameter $r_0$ in the topological sectors with charges $Q^{\rm i} \in \{0 \, , \, 1 \, , \, 2 \, , \, 4\}$.  
As can be seen from Table~\ref{TAB001}, there are substantial differences between results computed in different topological sectors going well beyond a $5 \sigma$ effect, in particular when comparing $Q^{\rm i} = 0$ and $Q^{\rm i} = 4$. 
%
%
This is also visible from fig.~\ref{FIG001}: while the static potentials for different $Q^{\rm i}$ sectors look rather similar (left), a zoomed plot shows a clear dependence on the topological charge (right). 
 
\begin{table}[htb] 
\begin{center} 
\begin{tabular}{|c||c|c|c||c|c||c|} 
\hline 
\vspace{-0.4cm} & & & & & & \\ 
$Q^{\rm i}$ & $\langle  W_{(1,1)} \rangle$ & $\langle  W_{(4,4)} \rangle$ & $\langle  W_{(7,7)} \rangle$ & $V(4)$ & $V(7)$ & $r_0/a$ \\ 
\vspace{-0.4cm} & & & & & & \\ 
\hline \hline 
0 & 0.94776(2) & 0.4090(6) & 0.0931(7) & 0.2177(3) & 0.3326(8) & 6.97(5) \\ \hline 
1 & 0.94770(3) & 0.4074(6) & 0.0911(8) & 0.2189(3) & 0.3364(9) & 6.84(4) \\ \hline 
2 & 0.94765(2) & 0.4057(6) & 0.0895(7) & 0.2201(3) & 0.3393(9) & 6.77(4) \\ \hline 
4 & 0.94743(2) & 0.4000(4) & 0.0832(5) & 0.2245(2) & 0.3515(7) & 6.48(3) \\ 
\hline 
\end{tabular} 
\end{center} 
 
\caption{\label{TAB001}Dependence of various observables related to the static potential on the topological sector defined by $Q^{\rm i}$. $\langle  W_{(R,T)} \rangle$ denotes smeared Wilson loop averages (extension $R \times T$), $V(R)$ the static potential at separation $R$ and $r_0$ the Sommer parameter (determinant ratio action, $\beta = 6.063$, lattice volume $16^4$).} 
\end{table} 
 
In a recent paper by the JLQCD Collaboration \cite{Aoki:2008tq} a similar analysis was carried out: $r_0$ was computed for the topological sectors $Q^{\rm i} \in \{0 \, , \, 2 \, , \, 4\}$, but no dependence on $Q^{\rm i}$ within statistical errors has been established. Note, however, that compared to our simulations the spatial extension of the lattice $L$ (in physical units) was larger by a factor of about $1.7$ and the temporal extension larger by a factor of $\approx 3.4$, yielding in total a factor of about $17$ for the lattice volume $V = L^3 \times T$. Assuming that the $Q^{\rm i}$ dependence of $r_0$ is of the same form as that derived for spectral quantities in \cite{brower:03,aoki:07} the absence of a Q-dependence is consistent, because the dominating finite volume correction, when fixing topology, is proportional to $1 / V$, which presumably would shrink the deviations seen here to the $1$ to $2 \sigma$ level. Note also that in \cite{Aoki:2008tq} dynamical quarks were used, which, in principle, could have an effect since topological fluctuations are suppressed. Moreover, the lattice spacing used in this reference was significantly larger ($0.12 \, \textrm{fm}$ compared to $0.07 \, \textrm{fm}$).


 
\section{Discussion and summary}

In this paper we were mainly concerned with the question, whether different definitions of the topological charge lead to consistent results, when using topology fixing gauge actions. To address this  
question, we employed two modifications of the Wilson plaquette action,  
the hyperbolic action, which implements L\"uscher's admissibility condition,  
and the determinant ratio action, which leads to a spectral gap. 
 
As a result we could establish that there is a correlation between the degree topology is fixed and the agreement of both topological charge definitions used herein in the sense that the better topology is fixed the better the agreement comes out. Let us accept for a second the index of the overlap operator at a given value of $m_0$ as a reference definition of the topological charge, which is not unreasonable given the conceptual cleanness. Then we find that the hyperbolic action is not able to fix topology completely, even for the smallest value of $\epsilon \approx 0.66$ 
reached in this paper. We argued that in fact with a value of $r_0 / a \approx 7$ it will not be possible to achieve such a complete fixing of the topological charge. On the other hand the determinant ratio action fixes topology at once for $r_0 / a \gtapprox 5.4$, 
i.e.\ already at rather coarse lattice spacings. Note that in \cite{Aoki:2008tq} it has been found that even at $r_0/a \approx 4.2$ no change of the topological charge could be observed. 
Of course, it is still possible that modifications of the overlap kernel operator such as changing the value of $m_0$ or smearing the gauge configurations may lead to different values of the topological charge $Q^{\rm i}$ (the index), a question that clearly deserves further studies. 
 
 
When fixing topology in a finite volume, as is necessary for
certain numerical simulations, a number of conceptual questions
arise. As a first step towards addressing such problems, we studied
the dependence of the static potential on the topological charge.
For our lattice size with linear extent of $L\approx 1.2 \, \textrm{fm}$ we
found a clear dependence of the static potential at distances
of about $R\approx 0.5 \, \textrm{fm}$, when computed in the topological
charge sectors $Q=0$ and $Q=4$.
It will be very interesting to see, whether this difference
vanishes when taking the infinite volume limit. It is our goal to
investigate the volume dependence of physical observables
in fixed topological sectors. This will also allow
us to test the predictions for the finite size effects given
in \cite{brower:03,aoki:07}.

\begin{figure}[t] 
\begin{center} 
{\includegraphics[width=0.50\linewidth,angle=90,angle=90,angle=90]{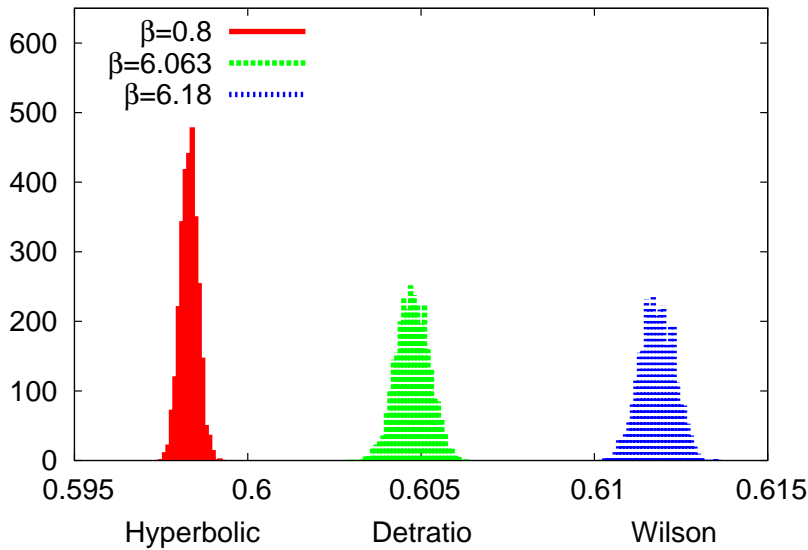}} 
\caption{The plaquette distributions for the Wilson, the hyperbolic ($\beta=0.8$) and 
the determinant ratio ($\beta=6.063$) action (roughly at a the same physical situation). Cf.\ also fig.~3 in \protect\cite{bietenholz:05}. 
} 
\label{fig:distributions} 
\end{center} 
\end{figure}

In this work we focused on {\em global} aspects of topology. It would also be interesting 
to investigate, how topology is locally realized in topology fixed gauge configurations.  
Moreover, we expect that the mechanism of topology fixing is rather different  
for the two actions we have used. This can be seen from fig.~\ref{fig:distributions}, 
where we show the plaquette distributions for the Wilson, the hyperbolic and  
the determinant ratio action at the same physical situation.  

As shown in more detail in fig.~3 of \cite{bietenholz:05} the hyperbolic action 
shows an asymmetric distribution for $1/\epsilon \gtapprox 1.52$,
 since small plaquette values are  
suppressed. In fact this suppression has been the original motivation  
for fixing topology. In contrast to that the distribution of the  
determinant ratio action is very similar to the distribution of the  
Wilson plaquette action. This can be made more quantitative by 
fitting Gaussian functions to these distributions with mean values $\mu$ and widths $\sigma$ as parameters.  
The results of these fits are listed in Table~\ref{table:disvalues} confirming that the value of $\sigma$ is almost smaller by a factor of two  
for the hyperbolic action. Nevertheless, the determinant ratio action  
fixes the reference topological charge completely. This result suggests that  
the suppression of small plaquette values  
is not a necessary  
condition for fixing topology. It would be interesting to investigate this in more detail. 
 
\begin{table}[htb] 
{\centering 
\begin{tabular}{|c|c|c|c|c|} 
\hline 
action &  $\beta$ & $1/\epsilon$ & $\mu$ & $\sigma$ \\ \hline \hline 
$S_{hyp}$ & 0.8 & 1.25 & 0.598360(9) & 0.000254(4) \\ \hline 
$S_{det}$ & 6.063 & - & 0.604808(9) & 0.000479(7) \\ \hline 
$S_{W}$ & 6.18 & - & 0.611857(9) & 0.000497(7) \\ 
\hline 
\end{tabular} 
\par} 
\caption{Results from Gaussian fits $\propto \exp^{-(x-\mu)^2 / 2 \sigma^2}$ to the plaquette distributions shown in fig.~\ref{fig:distributions}.} 
\label{table:disvalues} 
\end{table}


 
\section*{Acknowledgments} 
 
We thank Gunnar Bali, Gregorio Herdoiza, Michael M\"uller-Preussker, Silvia Necco, Luigi Scorzato and Andreas Sch\"afer for useful discussions and in particular Urs Wenger for helpful cross checks of static potential computations.
 
This work has been supported in part by the DFG Sonderforschungsbereich/Transregio SFB/TR9-03 and by DFG BR 2872/4-1. 

\bibliographystyle{unsrt} 
\bibliography{draft}

\end{document}